\documentclass[onecolumn,11pt,peerreview,draftclsnofoot]{IEEEtran}

\usepackage{graphicx}
\usepackage{epsfig}
\usepackage{dcolumn}
\usepackage{bm}
\usepackage{amsmath}
\usepackage{amssymb}

\begin{document}
%
\title{Memcapacitive neural networks}
%
%
%

\author{Yuriy~V.~Pershin and Massimiliano~Di~Ventra
\thanks{Y. V. Pershin is with the Department of Physics
and Astronomy and USC Nanocenter, University of South Carolina, \newline
Columbia, SC, 29208 \newline e-mail: pershin@physics.sc.edu.}
\thanks{M. Di Ventra is with the Department
of Physics, University of California, San Diego, La Jolla,
California 92093-0319 \newline e-mail: diventra@physics.ucsd.edu.}

}

%
%


\maketitle

\begin{abstract}
We show that memcapacitive (memory capacitive) systems can be used as synapses in
artificial neural networks.  As an example of our approach,
we discuss the architecture of an integrate-and-fire neural network based on memcapacitive synapses.
Moreover, we demonstrate that the spike-timing-dependent plasticity can be
simply realized with some of these devices. Memcapacitive synapses are a low-energy alternative
to memristive synapses for neuromorphic computation.
\end{abstract}


%
\IEEEpeerreviewmaketitle

\IEEEPARstart{E}{lectronic} devices with memory such as memristive \cite{chua76a}, memcapacitive and meminductive systems \cite{diventra09a} are promising components for unconvential computing applications because of their ability to store and process information at the same space location \cite{diventra13a}. Moreover, if these devices are used as memory and computing elements in, e.g., neural networks, high-integration density and low-power consumption can be easily achieved. So far, only memristive devices have been considered as electronic synapses in artificial neural networks \cite{snider08a,jo10a,pershin10c,pershin12a}. In this article,
we show instead that memcapacitive systems could play a similar role, thus offering the benefit of low power dissipation \cite{diventra09a} and, in some instances, full compatibility with CMOS technology \cite{traversa13b}, an added benefit for electronic realizations of "smart electronics".

According to their definition \cite{diventra09a}, voltage-controlled memcapacitive systems are described by the equations
\begin{eqnarray}
V_C(t)&=&C^{-1}\left(x,q,t \right)q(t) \label{Ceq1} \\
\dot{x}&=&f\left( x,q,t\right) \label{Ceq2}
\end{eqnarray}
where $q(t)$ is the charge on the capacitor at time $t$, $V_C(t)$
is the applied voltage, $C$ is the {\it memcapacitance}, which
depends on the state of the system and can vary in time, $x$ is a
set of $n$ state variables describing the internal state of the system, and $f$ is a continuous
$n$-dimensional vector function. It is currently well established that in
biological neural networks the synapse strength encodes memories \cite{Cowan01a}. In electronic circuits, the memory feature of
memcapacitive systems (provided by their internal states characterized by $x$) can play a similar role.

\begin{figure} [b]
 \begin{center}
\includegraphics[width=7.5cm]{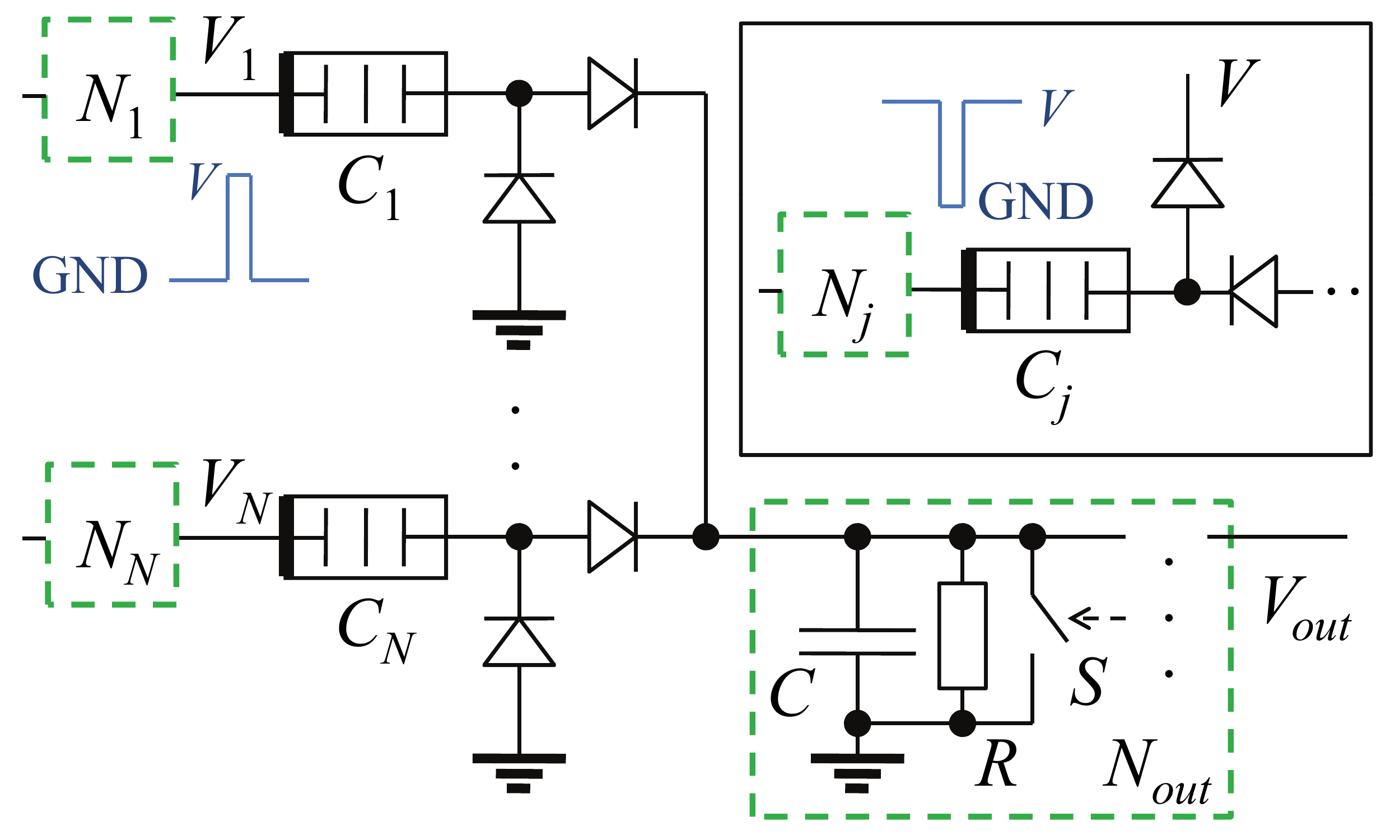}
\caption{Memcapacitive synapses in an integrate-and-fire neural network. $N$ input electronic neurons $N_1$-$N_N$ are connected to the output neuron $N_{out}$ using memcapacitive synapses. The inset: schematics of the inhibitory synapse. \label{fig1}}
 \end{center}
\end{figure}

In figure \ref{fig1} we show an example of memcapacitive neural network. In this leaky integrate-and-fire network, $N$ input neurons are connected to the $RC$ block of the output neuron with the help of memcapacitive synapses $C_1-C_N$. Each memcapacitive synapse contains a memcapacitive system and two diodes. It is assumed that the switching of memcapacitive system involves a voltage threshold, which is above voltage pulse amplitudes involved in this network. Subjected to a voltage pulse from the $i$-th input neuron, the memcapacitive system $C_i$ charges the integrating capacitor $C$ in proportion to its capacitance $C_i$. As soon as the voltage across $C$ reaches the threshold of $N_{out}$, $N_{out}$ fires a forward voltage pulse and uses the controllable switch $S$ to reset $C$. The diodes connected to the ground discharge $C_1-C_N$ after the pulse disappearance, and can be replaced by resistors.

\begin{figure}
 \begin{center}
\includegraphics[width=7cm]{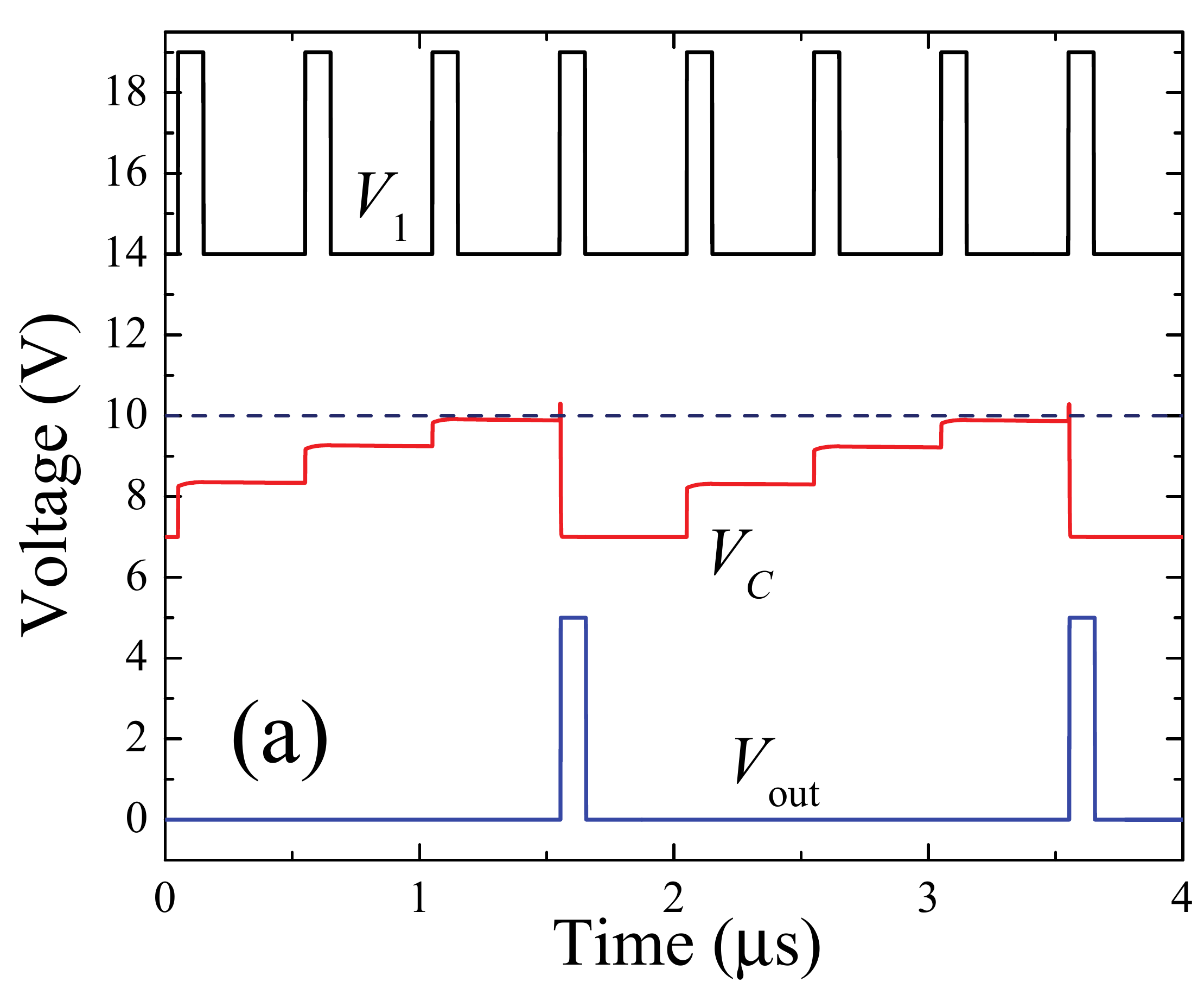}
\vspace{0.3cm}
\includegraphics[width=7cm]{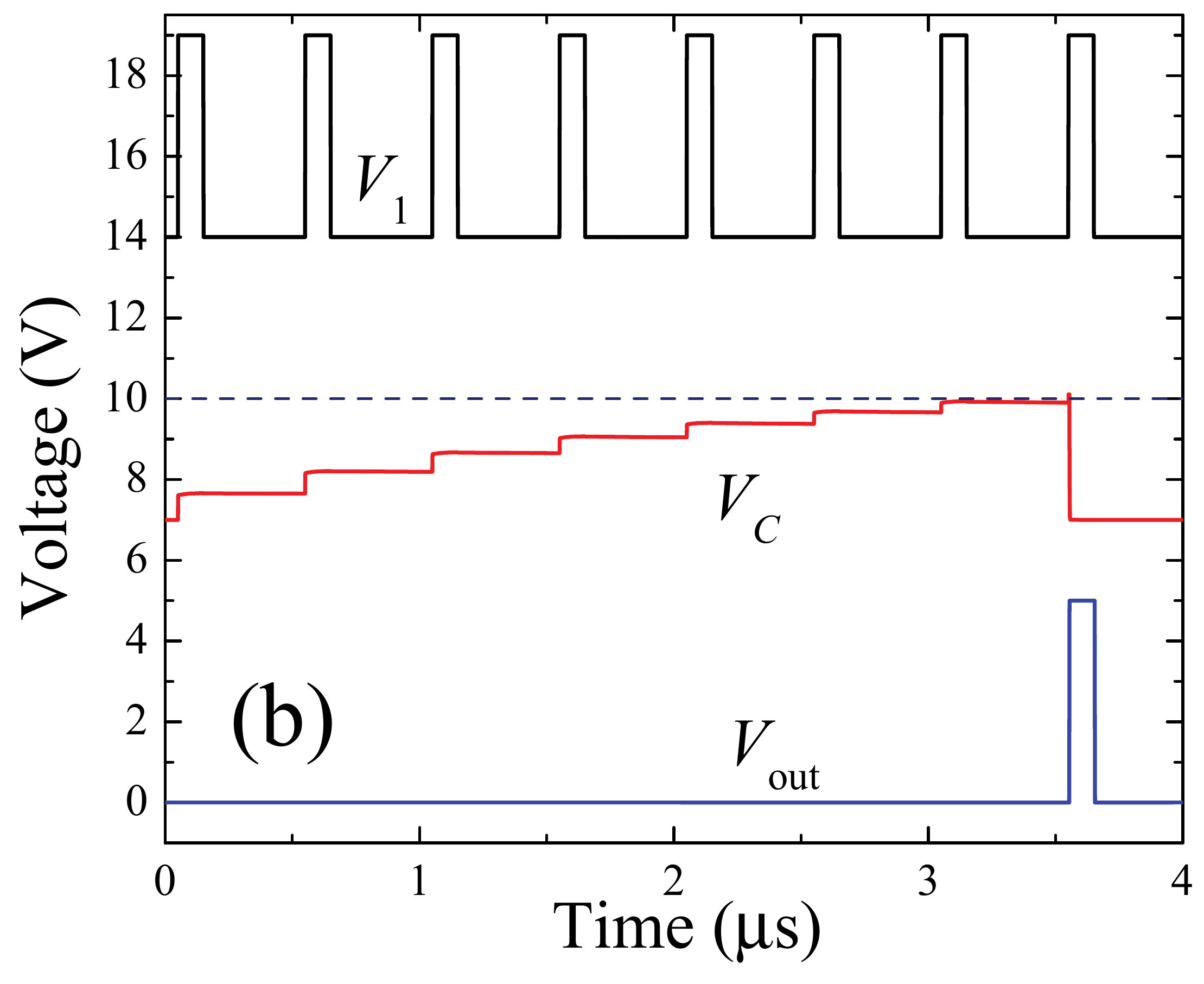}
\caption{Simulation of the integrate-and-fire memcapacitive network (Fig. \ref{fig1}) with
only one spiking neuron. The regular 5V 100ns spikes of the input neuron $N_1$ trigger the output neuron $N_{out}$ (with 3V voltage threshold shown by the horizontal dashed line) at different frequencies depending on the strength of memcapacitive synapse ($C_1$=2nF in (a) and $0.8$nF in (b)). This plot was obtained with $C=5$nF, $R=10$k$\Omega$, and BAT54 diode model. Here, $V_C$ is the voltage across $C$, and the lines are shifted by 7V for clarity. \label{fig2}}
 \end{center}
  \vspace{-10pt}
\end{figure}

Fig. \ref{fig2} presents simulations of the integrate-and-fire network in the LTspice environment assuming the firing of only one input neuron. Here, we compare the circuit response subjected to the same input pulse sequence (periodic firing of $N_1$) at different values of corresponding synaptic connection (memcapacitance $C_1$). The stronger synaptic connection (Fig. \ref{fig2}(a)) results in faster charging of the integrating capacitor $C$ and higher rate of the output neuron $N_{out}$ firing. This result is compatible with operation of excitatory synapses. In order to model inhibitory synapses, one can use the synaptic connection similar to that shown in Fig. \ref{fig1} with diodes connected with opposite polarity to the integrating capacitor and power supply voltage $V$ (instead of the ground). The inset in Fig. \ref{fig1} shows the schematics of the inhibitory synapse explicitly.  Moreover, the inhibitory synapse should be driven by inverted pulses.

To evaluate the strengths and weaknesses of using memcapacitive systems as synapses, we compare the energy dissipation in memcapacitive and memristive neural networks. Indeed, the circuit depicted in Fig. \ref{fig1} can also operate with memristive synapses replacing the memcapacitive ones. Let us then estimate the amount of energy lost in both cases. For this purpose, we consider the situation when a single voltage pulse fired by $N_1$ charges $C$ by a small amount of charge $q$ from $V_C=0$. In the case of memcapacitive network, the dissipated energy is the energy stored in $C_1$ due to the pulse, namely, $U_{C}=qV_1/2\approx qV/2$ if $C_1\ll C$. In the case of memristive network, $U_R\approx \Delta t V I\approx qV$. Consequently, in this application, the memcapacitive synapses are two times more energy efficient. However, the memristive networks require less components as the diodes connected to the ground in Fig. \ref{fig1} are not required when memristive synapses are not used.

Let us now consider a realization of the spike-timing-dependent plasticity (STDP) with memcapacitive synapses. For this purpose, we consider
a bipolar memcapacitive system with threshold \cite{diventra09a,biolek13a}, which is also suitable for the integrate-and-fire network considered above.
In biological neural networks, when a postsynaptic signal reaches the synapse before the action potential of the presynaptic neuron, the
synapse shows long-term depression (LTD), namely its
strength decreases (weaker connection between the neurons) depending on the time difference between the postsynaptic and presynaptic signals. Conversely, when the
postsynaptic action potential reaches the synapse after the
presynaptic action potential, the synapse undergoes a
longtime potentiation (LTP), namely the signal transmission between the two neurons increases in proportion to
the time difference between the presynaptic and postsynaptic signals.

\begin{figure}
 \begin{center}
\includegraphics[width=7cm]{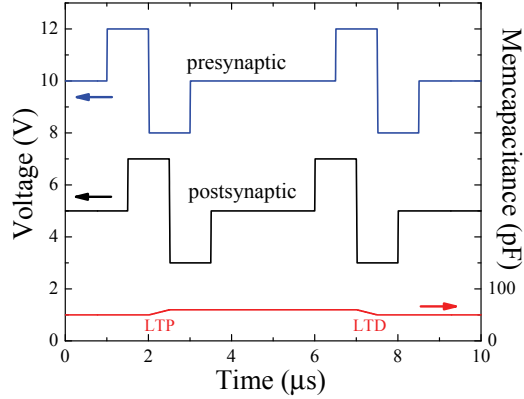}
\caption{STDP with memcapacitive synapses. This plot was obtained using a model of bipolar memcapacitive system with threshold \cite{biolek13a} with $V_\textnormal{t}=3$V, $C_\textnormal{low}=10$pF, $C_\textnormal{high}=100$pF, $C(t=0)=50$pF, $\beta=2\times 10^{-5}$F/(V$\cdot$s). Here, the lines are shifted by 5V for clarity.\label{fig3}}
 \end{center}
 \vspace{-10pt}
\end{figure}

In electronic circuits, STDP can be implemented using double voltage pulses as shown in Fig. \ref{fig3} (see also Ref. \cite{pershin12a}).
In this case, the pulse overlap provides opposite voltage polarities (across the synapse) depending on timing of presynaptic and postsynaptic pulses.
Using a SPICE model of memcapacitive system with threshold \cite{biolek13a} we simulate the dynamics of a memcapacitive synapse subjected to LTP and LTD pulses.
The bottom line in Fig. \ref{fig3} clearly shows the corresponding increase and decrease of the synaptic strength (memcapacitance).

In conclusion, we have presented an alternative to simulate synaptic behavior that uses memcapacitive
systems instead of memristive systems. The corresponding memcapacitive neural networks can operate at
low energy consumption, and in some cases they are compatible with CMOS technology making them
promising candidates for neuromorphic computing.

This work has been partially supported by NSF grant ECCS-1202383, the Center for Magnetic Recording Research at UCSD, and Burroughs Wellcome Fund 2013 Collaborative Research Travel Grant.

\bibliographystyle{IEEEtran}
\bibliography{IEEEabrv,maze}

\end{document}